\documentclass[preprint,tightenlines, nofootinbib,aps,prc]{revtex4-1} 

\usepackage{graphicx}% Include figure files
\usepackage{bm}% bold math
\usepackage{url,multirow,xcolor}
\begin{document}

\title{ Determination of Strength of Isoscalar Pairing Interaction 
by a Mathematical Identity in QRPA }\thanks{Presented in Zakopane Conference on Nuclear Physics ``Extreme of the Nuclear Landscape", August 26$-$September 2, 2018, Zakopane, Poland.}

\author{J.\ Terasaki}
\affiliation{ Institute of Experimental and Applied Physics, Czech Technical University in Prague, Horsk\'a 3a/22, 128 00, Prague 2, Czech Republic \\ }

\begin{abstract} 
I propose a new method to determine the strength of the isoscalar proton-neutron pairing interaction by a  mathematical identity derived in the quasiparticle random-phase approximation. This method is applied for a few nuclei possibly having the neutrinoless double-$\beta$ decay. Reduced half-life, the theoretical quantity necessary for determining the effective neutrino mass, is calculated for $^{48}$Ca. 
\end{abstract}

%\pacs{}
%\keywords{QRPA, deformed, overlap}
%\author{overlap9.tex, Aug.~15, 2012}
%
\maketitle
\section{Introduction \label{sec:introduction}}
It has been discussed that the proton-neutron pairing interaction was important in the calculation of the nuclear matrix elements of the neutrinoless double-$\beta$ decay by the quasiparticle random-phase approximation (QRPA), e.g., \cite{Eng17}. Assuming the isospin symmetry, one can use the average of the proton-proton and neutron-neutron pairing-interaction strengths for the isovector proton-neutron pairing-interaction strength. As for the isoscalar pairing-interaction strength, it has been determined indirectly by reproducing experimental data not reflecting the proton-neutron pairing correlations as strongly as the pairing gap, e.g., \cite{Sim13}. This approach was unavoidable because the proton-neutron pairing gap is not established in experimental data. 

In this paper, I propose a new method to determine the strength of the isoscalar pairing interaction by calculating a transition matrix element of double-charge-change operator.  This method is presented in Sec.~2, and the calculated result of the reduced half-life of the neutrinoless double-$\beta$ decay is shown for $^{48}$Ca in Sec.~3. The reduced half-life is the theoretical quantity necessary for determining the effective neutrino mass, and the reliable calculation of this physical quantity is a goal of the theoretical study of the neutrinoless double-$\beta$ decay. Section 4 is summary. The detail of this method is discussed in Ref.~\cite{Ter16}.

\section{The new idea}
Let me consider the calculation of the transition matrix element
\begin{eqnarray}
M^{(0\nu)} = \langle 0^+_\textrm{\scriptsize{f}} | T^{(0\nu)} | 0^+_\textrm{\scriptsize{i}} \rangle , 
\end{eqnarray}
where $|0^+_\textrm{\scriptsize{i}}\rangle$ and $|0^+_\textrm{\scriptsize{f}}\rangle$ are the ground states of even-even nuclei with different proton number $Z$ and neutron number $N$ as $(Z,N)$ for $|0^+_\textrm{\scriptsize{i}}\rangle$ and $(Z+2,N-2)$ for $|0^+_\textrm{\scriptsize{f}}\rangle$. The transition operator is a double-charge-change operator 
\begin{eqnarray}
T^{(0\nu)} = \sum_{pp^\prime nn^\prime} \langle pp^\prime | V(\bm{r}) | nn^\prime \rangle 
c^\dagger_{p^\prime} c_{n^\prime} c^\dagger_p c_n .
\end{eqnarray}
Symbols $p, p^\prime$ and $n,n^\prime$ denote the proton and neutron, respectively, and $c^\dagger_i$ and $c_i$ ($i$: single-particle state) are the creation and annihilation operators, respectively. $V(\bm{r})$ is a two-body potential including an appropriate isospin operator. 

$M^{(0\nu)}$ can be calculated in two ways in the QRPA. One is to use 
\begin{eqnarray}
M^{(0\nu)} = \sum_{pp^\prime nn^\prime} \langle pp^\prime | V(\bm{r}) | nn^\prime \rangle 
\sum_{ b^\textrm{\scriptsize{pn}}_\textrm{\scriptsize{f}}:\textrm{\scriptsize{pnQRPA}} }
\sum_{ b^\textrm{\scriptsize{pn}}_\textrm{\scriptsize{i}}:\textrm{\scriptsize{pnQRPA}} }
\langle 0^+_\textrm{\scriptsize{f}} | c^\dagger_{p^\prime} c_{n^\prime} | b^\textrm{\scriptsize{pn}}_\textrm{\scriptsize{f}} \rangle 
\langle b^\textrm{\scriptsize{pn}}_\textrm{\scriptsize{f}} | b^\textrm{\scriptsize{pn}}_\textrm{\scriptsize{i}} \rangle \langle b^\textrm{\scriptsize{pn}}_\textrm{\scriptsize{i}} | c^\dagger_p c_n | 0^+_\textrm{\scriptsize{i}} \rangle , \label{eq:M0v-1}
\end{eqnarray}
where $|b^\textrm{\scriptsize{pn}}_\textrm{\scriptsize{i}}\rangle$'s and $|b^\textrm{\scriptsize{pn}}_\textrm{\scriptsize{f}}\rangle$'s  are the states of the proton-neutron QRPA (pnQRPA), and the formers (latters) are obtained on the basis of the initial (final) Hartree-Fock-Bogoliubov ground state. Another way is to use 
\begin{eqnarray}
M^{(0\nu)} = \sum_{pp^\prime nn^\prime} \langle pp^\prime | V(\bm{r}) | nn^\prime \rangle 
\sum_{ b^\textrm{\scriptsize{lp}}_\textrm{\scriptsize{f}}:\textrm{\scriptsize{lpQRPA}} }
\sum_{ b^\textrm{\scriptsize{lp}}_\textrm{\scriptsize{i}}:\textrm{\scriptsize{lpQRPA}} }
\langle 0^+_\textrm{\scriptsize{f}} | c_n c_{n^\prime} | b^\textrm{\scriptsize{lp}}_\textrm{\scriptsize{f}} \rangle 
\langle b^\textrm{\scriptsize{lp}}_\textrm{\scriptsize{f}} | b^\textrm{\scriptsize{lp}}_\textrm{\scriptsize{i}} \rangle \langle b^\textrm{\scriptsize{lp}}_\textrm{\scriptsize{i}} | c^\dagger_{p^\prime} c^\dagger_p | 0^+_\textrm{\scriptsize{i}} \rangle , \label{eq:M0v-2}
\end{eqnarray}
where $|b^\textrm{\scriptsize{lp}}_\textrm{\scriptsize{i}}\rangle$'s and $|b^\textrm{\scriptsize{lp}}_\textrm{\scriptsize{f}}\rangle$'s  are the states of the like-particle QRPA (lpQRPA), respectively. 
The ground states of Eqs.~(\ref{eq:M0v-1}) and (\ref{eq:M0v-2}) are those of the pnQRPA and lpQRPA, respectively.\footnote{For the mathematical completeness, it is necessary to consider the product states of the lp- and pn- QRPA ground states; see Ref.~\cite{Ter16} for this extension. The simplification in this paper does not affect the new method.}
For the calculation of overlap 
$\langle b^\textrm{\scriptsize{pn}}_\textrm{\scriptsize{f}} | b^\textrm{\scriptsize{pn}}_\textrm{\scriptsize{i}} \rangle$ and 
$\langle b^\textrm{\scriptsize{lp}}_\textrm{\scriptsize{f}} | b^\textrm{\scriptsize{lp}}_\textrm{\scriptsize{i}} \rangle$, 
see Refs.~\cite{Ter12, Ter13}. 
The two variants of the QRPA do not satisfy the equality of Eqs.~(\ref{eq:M0v-1}) and (\ref{eq:M0v-2}) for {\it arbitrary} interactions. 
It is, however, sufficient, if the interaction used for the calculation satisfies the equality. Thus, the equality of the two equations is a constraint on the effective interactions for the QRPA, and the strength of the isoscalar pairing interaction can be determined by these equations under the presumption that the other interactions are established. 
%\subsection{Subsection}
%The text...

\section{Application to neutrinoless double-$\beta$ decay}
I have so far applied the above new idea to four candidates of the neutrinoless double-$\beta$ decay: $^{48}$Ca-$^{48}$Ti \cite{Ter18}, $^{130}$Te-$^{130}$Xe, $^{136}$Xe-$^{136}$Ba, and $^{150}$Nd-$^{150}$Sm \cite{Ter16,Ter15} (calculations for other candidates are in progress). 
The neutrino potential arising from the neutrino-exchange interaction between nucleons was used for $V(\bm{r})$, e.g., \cite{Doi85}. 
For the particle-hole interactions, the Skyrme (SkM$^\ast$ \cite{Bar82}) was used. All the pairing interactions are of the contact volume type, and the strengths of the like-particle pairing interactions were determined from the experimental odd-even mass differences as usual (the three-point formula \cite{Boh69}). The strengths of the isovector proton-neutron pairing interaction was determined from these strengths as mentioned in Sec~\ref{sec:introduction}. It turns out that the strengths of the isoscalar pairing interaction of those nuclei are in the range 
between $-50.0$ and $-197.4$ MeV$\,$fm$^3$. Those of the like-particle pairing interactions are in the range between $-176.4$ and $-224.5$ MeV$\;$fm$^3$. Thus, the strengths by the new method are sometimes significantly smaller than those of the like-particle pairing interactions but not larger. 

 Calculation was performed for the reduced half-life $R^{(0\nu)}_{1/2}$ to the neutrinoless double-$\beta$ decay. It is defined by 
\begin{eqnarray}
T^{(0\nu)}_{1/2} = R^{(0\nu)}_{1/2} \langle m_\nu \rangle^{-2} ,
\end{eqnarray}
where $T^{(0\nu)}_{1/2}$ is the half-life to the neutrinoless double-$\beta$ decay, and $\langle m_\nu \rangle$ is the effective neutrino mass defined by a transformation from the three neutrino masses, e.g., \cite{Doi85}. $R^{(0\nu)}_{1/2}$ is obtained from the nuclear matrix element calculated from the nuclear wave functions and the phase-space factor arising from the emitted electrons, e.g., \cite{Sim13,Ter16}. If $T^{(0\nu)}_{1/2}$ is obtained experimentally, $\langle m_\nu \rangle$ can be determined using the theoretical $R^{(0\nu)}_{1/2}$; this determination is a major goal of the study of the neutrinoless double-$\beta$ decay. This decay not yet observed occurs, if and only if the neutrino is a Majorana particle \cite{Fur39}. Thus, the primary purpose of the experiments is to clarify this nature of the neutrino. 

\begin{figure}[t]
\centerline{%
\includegraphics[width=10.0cm]{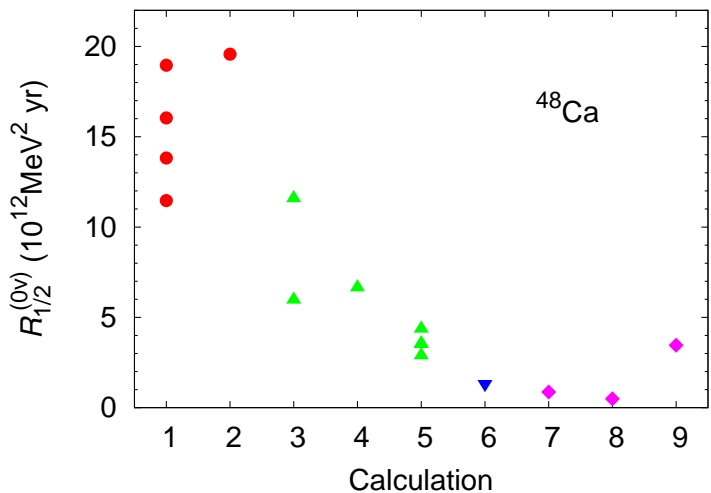}}
\caption{$R^{(0\nu)}_{1/2}$ calculated for $^{48}$Ca-$^{48}$Ti by different groups using several methods indicated by the calculations  1$-$9. The method of 1 and 2 is the QRPA, and the latter is my result. That of 3$-$5 is the shell model, that of 6 is the interacting boson model-2, and that of  7$-$9 is the generator-coordinate method. The effective axial-vector current coupling is not unified (that is the reason why $R^{(0\nu)}_{1/2}$ is shown); see the references below. The references and corresponding calculation numbers are as follows: \cite{Sim13} (1), \cite{Ter18} (2), \cite{Men09} (3), \cite{Hor16} (4), \cite{Iwa16} (5), \cite{Bar15} (6), \cite{Vaq13} (7), \cite{Yao15} (8), and \cite{Jia17} (9).}
\label{fig:reduced_half-life}
\end{figure}

Figure \ref{fig:reduced_half-life} shows the calculated $R^{(0\nu)}_{1/2}$ for $^{48}$Ca including those of other groups. A discussion is possible assuming a value of $\langle m_\nu \rangle$, for example 10 meV. With this value and my value of $R^{(0\nu)}_{1/2}$ a predicted $T^{(0\nu)}_{1/2}$ is obtained to be 2$\times$10$^{29}$ years. This value can be compared to the age of the universe  $\approx$ 13.8$\times$10$^9$ years \cite{Pla16}. It is seen that the neutrinoless double-$\beta$ decay is an extremely rare decay. 

\section{Summary}
I have proposed a new method to determine the strength of the isoscalar pairing interaction by using the two expressions of the double-charge-change transition matrix element, which are identical by the appropriate effective interaction. This method is useful particularly because the proton-neutron pairing gap is not established. The motivation for developing this method is to eliminate an uncertainty from the calculation of the reduced half-life of the neutrinoless double-$\beta$ decay using the QRPA. The new method has been applied to four decay instances, and in this paper $R^{(0\nu)}_{1/2}$ for $^{48}$Ca was shown, which indicates that the half-life to the neutrinoless double-$\beta$ decay is extremely long. 

%\begin{acknowledgments} 
\vspace{15pt}
%\begin{center} 
%\bf Acknowledgments
%\end{center}
%\vspace{3pt}
The numerical calculations of this paper were performed by 
the K computer at RIKEN Center for Computational Science, through the program of High Performance Computing Infrastructure in 2016 (hp160052)  and 2017$-$2018 (hp170288). Computer Coma at Center for Computational Sciences, University of Tsukuba was also used through
Multidisciplinary Cooperative Research Program of this center in 2016 (TKBNDFT) and 2017 (DBTHEORY). 
Further computer Oakforest-PACS at Joint Center for Advanced High Performance Computing was used through the above program of Center for Computational Sciences, University of Tsukuba in 2018 (xg18i006). 
This study is supported by 
European Regional Development Fund-Project ``Engineering applications of microworld physics" (No. CZ.02.1.01/0.0/0.0/16\_019/0000766).
%Underground laboratory LSM - Czech participation to European-level research infrastructure cz.02.1.01$/$0.0$/$0.0$/$16\_013$/$0001733. 
%\end{acknowledgments}

 \end{document}